\documentclass[twocolumn,showpacs,preprintnumbers,amsmath,amssymb]{revtex4}

\usepackage{graphicx}
\usepackage{dcolumn}
\usepackage{bm}
\begin{document} 

\title{Single particle spectra of charge transfer insulators by cluster
perturbation theory - the correlated band structure of NiO}
\author{R. Eder$^1$, A. Dorneich$^2$ and H. Winter$^1$}
\affiliation{$^1$ Forschungszentrum Karlsruhe, Institut f\"ur
  Festk\"orperphysik, 76021 Karlsruhe, Germany\\
$^2$ IBM Deutschland Entwicklung GmbH, 71032 B\"oblingen, Germany}
\date{\today}
\begin{abstract}
We propose a many-body method for band-structure calculations
in strongly correlated electron systems and apply it to NiO. 
The method may be viewed as a translationally invariant version of 
the cluster method of Fujimori and Minami. Thereby the Coulomb
interaction within the $d$-shells is treated by exact diagonalization 
and the $d$-shells then are coupled to a solid by an extension of the
cluster perturbation theory (CPT) due to Senechal et al. The method
is computationally no more demanding than a conventional band
structure calculation and for NiO we find good agreement between the 
calculated single particle spectral function and the experimentally 
measured band structure. 
\end{abstract} 
\maketitle
\section{Introduction}
Band structure calculations based on the single particle picture
have enjoyed considerable success in solid state theory.
Single particle picture here is meant to imply that the ground state
is obtained by filling up according to the Pauli principle
the energy levels calculated for a single electron
in an `effective potential'.
The effective potential thereby is usually constructed
within the framework of the local density (LDA)
or local spin density approximation (LSDA) to
density functional theory (DFT)\cite{KohnSham} and 
despite the well-known fact that the eigenvalues of the Kohn-Sham
equations should not be identified with the single-particle
excitation energies of a system, the resulting band structures
often give an almost quantitative description of angle-resolved 
photoemission spectroscopy (ARPES). \\
However, there are also some classes of solids which defy such a
description, most notably transition metal compounds with partially 
filled $d$ and $f$-shells and strong Coulomb interaction between the 
electrons in these. A frequently cited example is NiO, where LSDA band
structure calculations correctly predict an antiferromagnetic and insulating
ground state, but only a small `Slater gap' of a fraction of
an eV\cite{Terakura}, whereas experimentally NiO is an insulator with
a bandgap of 4.3 eV\cite{SawatzkyAllen} and stays so even above
the magnetic ordering temperature.  While DFT thus gives
reasonable answers within its domain of validity - namely ground
state properties - the non-correspondence between 
the Kohn-Sham eigenvalues and the single-particle excitation
energies of the solid obviously has to be taken literal for this 
compound (if the band gap is not read off from the
LSDA band structure but expressed as the difference
of ground state energies it is in fact
possible to calculate it within the framework
of DFT, as shown by Norman and Freeman\cite{NormanFreeman}). 
It is generally believed that
the reason for the discrepancy is the strong Coulomb interaction
between the electrons in the Ni $3d$-shell, which
leads to a substantial energy splitting between
$d^n$ configurations with different $n$.
This leads to a very pronounced `pinning' of the $d$-shell occupation
number $n$, in the case of NiO to the value $n=8$.
Final states for photoemission or inverse photoemission then
correspond to a single $d$-shell being in either a $d^7$, a $d^8$
or a $d^9$ configuration, in each case with very small
admixture of configurations with other
$n$. The corresponding `defect' then may be thought of
propagating through
the crystal with definite $\bf{k}$. 
This pinning of the electron number in both initial and
final states cannot be reproduced by a wave function which
takes the form of a simple Slater-determinant - such as
the ground state deduced from the Kohn-Sham equations.
The situation is improved somewhat in the
self-interaction corrected version of DFT\cite{SvaneGunnarsson,Szotek},
which renders a certain fraction
of the $d$-orbitals completely localized, so that
their occupation number in fact does become pinned - for the 
remaining delocalized $d$-orbitals, however, the problem remains.
Another way to achive the pinning of the $d$-shell
occupancy is the use of an orbital-dependent potential
in the framework of the so called LDA+U method\cite{Anisimov,Czyzyk}.
Speaking about gap values the calculations based
on the GW-approximation\cite{Aryasetiawan,Massida} 
also need to be mentioned - these give {\em ab initio} gap values
which are in good agreement with  experiment
but do not seem to reproduce the high energy `satellite'
in the photoemisison spectrum.\\
In addition to the pinning of the $d$-shell occupancy,
the multiplet structure of the metal ion poses a problem
for single-particle theories as well. It is quite well established that
the multiplet structure (appropriately modified by
the crystal field splitting) of the isolated metal ion persists in 
the solids. Clear evidence for this point of view comes from the
fact, that  angle integrated valence 
band photoemission spectra\cite{FujimoriMinami} as well as
X-ray absorption spectra\cite{Elp} of many transition
metal compounds can be reproduced in remarkable detail by 
configuration interaction calculations solving 
exactly the problem of a single $d$-shell hybridizing with a `cage' of 
ligands. In these calculations it is crucial, however,
that the intra-shell Coulomb repulsion is treated in full detail.
While the cluster method is spectacularly successful
for angle-integrated quantities its `impurity' character
unfortunately makes it impossible to extract the dispersion relations
of $\bm{k}$-resolved single particle excitations.\\
Actual dispersion relations in the presence of strong
Coulomb interaction were first studied by Hubbard\cite{Hubbard},
thereby taking an entirely different point of view as compared to 
the single particle picture on which conventional band structure 
calculations are based. Thereby the $d$-shells first are considered
as isolated, and their affinity and ionization spectra
obtained, thereby treating the Coulomb repulsion
exactly. In his famous papers Hubbard used a much simplified model, 
where the orbital degeneracy of the $d$-level was neglected whence
ionization and affinity spectrum of the `half-filled' $d$-shell
collapse to single peak each, with the two peaks separated by the
Coulomb energy $U$. Upon coupling the individual atoms to the solid, the
ionization and affinity states of the individual
atoms then are systematically broadened to form the two
`Hubbard bands'. The coupling to the solid was achieved originally by the
famous Hubbard I approximation, but in fact
this may be interpreted as a particularly simple form of the 
cluster perturbation theory (CPT), proposed by 
Senechal {\em et al.}\cite{Senechaletal},
where the individual `clusters' consist of just a single $d$-shell.
This suggests immediately to relax Hubbard's simplifications and 
take into account the full complexity of a transition metal oxide
including the orbital degeneracy of the $d$-shell, the full
Coulomb interaction between $d$-electrons in these and the sublattice of
ligands. This is essentially the purpose of the present manuscript.\\
An important complication is due to the sublattice of ligands.
It has been shown by Fujimori and Minami\cite{FujimoriMinami} that 
in discarding altogether 
the sublattice of ligands (in the case of NiO: the oxygen atoms),
Hubbard actually went 
one step to far in his simplification of the model.
Namely Fujimori and Minami showed
that the top of the valence band in NiO is composed of states,
where a hole is predominantly in an oxygen atom, but somehow
`associated' with an $n$-conserving excitation of a
neighboring $d$-shell - i.e. a magnon or a $d-d$ exciton.
This type of state might be viewed as generalization of a Zhang-Rice 
singlet\cite{ZRS} in the CuO$_2$ planes of cuprate superconductors.
It was then found by Zaanen, Sawatzky and Allen\cite{ZSA}
that there is a crossover between
this so-called charge-transfer insulator and a more conventional
Mott-Hubbard insulator as a function of two key parameters,
the Coulomb repulsion $U$ between electrons in the $d$-shell and
the charge transfer energy $\Delta$, which are defined as
\begin{eqnarray*}
E(d^n\rightarrow d^{n+1}\underline{L}) &=& U-\Delta, \nonumber \\
E(d^n\underline{L} \rightarrow d^{n-1}) &=& \Delta. 
\end{eqnarray*}
Strong experimental support for the picture proposed by Fujimori and Minami
is provided by the resonance behavior of the photoemission intensity as seen in
photoemission with photon energies near the
$2p\rightarrow 3d$ absoption threshold\cite{Ohetal,Tjernberg,Tjeng}.\\
Adopting this point of view and using a simplified
`Kondo-Heisenberg' model,
in which the charge degrees of freedom on Ni where projected out,
Bala {\em et al.}\cite{BalaOlesZaanen_I} then obtained
dispersion relations of quasiparticles in NiO which in fact
do contain the key feature seen
in ARPES\cite{Kuhlenbeck,Shen_long}:
the coexistence of strongly dispersive oxygen bands on one hand
and a complex of practically dispersionless (i.e.: massively 
renormalized) bands which form the top of the valence band 
on the other hand. \\
In the present theory no reduction of the Hamiltonian to a
t-J-type model is performed. Rather we use the same basic idea
as in the treatment of the Kondo lattice in Ref. \cite{Oana}:
the system is divided into sub-units which are treated
exactly and the hybridization between the sub-units is
treated approximately. To do so, we define the ground state for
vanishing hybridization as the `vacuum state' and treat the
charge fluctuations created by the hybridization as `effective
Fermions', for which an approximate Hamiltonian can be derived and solved.
It has been shown in Ref. \cite{Ansgar} that the Hubbard-I
approximation for the single-band Hubbard model
can be re-derived in this fashion if the
sub-units are taken to be only a single site - including more
complex `composite particles' which extend over several unit
cells then improves the agreement with numerical results.
The generalized Zhang-Rice singlets discussed above
may be viewed as such composite particles.
It is shown in Appendix I of Ref. \cite{Marc} that this treatment
is in fact equivalent to the original cluster perturbation theory
of Senechal {\em et al.}\cite{Senechaletal} provided
the sub-units into which the system is divided are non-overlapping. 
This last requirement poses a substantial
problem for transition metal oxides,
because the rocksalt lattice of NiO cannot be easily divided into
non-overlapping sub-units which are still amenable
to exact diagonalization without artificially breaking
a symmetry of the lattice (which would lead to
artificial symmetry breaking in the band structure).
We therefore need to adjust the concept of cluster perturbation theory
to this situation - which also is an objective of the present
work. The remainder of the paper is organized as follows: in section
II we discuss a simplified 1D model, in section III
we present the general theory, in section IV we apply the
theory to the 1D model and compare the obtained single particle
spectra with result from exact diagonalization, in
section IV we discuss the ARPES spectra of NiO and section V
gives the conclusions.
\section{A simplified model}
\begin{figure}
\includegraphics[width=\columnwidth]{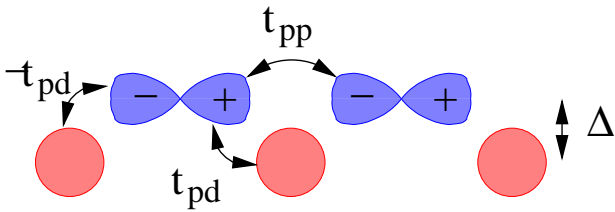}
\caption{\label{fig3} Schematic representation of the Hamiltonian
(\ref{model}) and its parameters.}
\end{figure}

\begin{figure}
\includegraphics[width=\columnwidth]{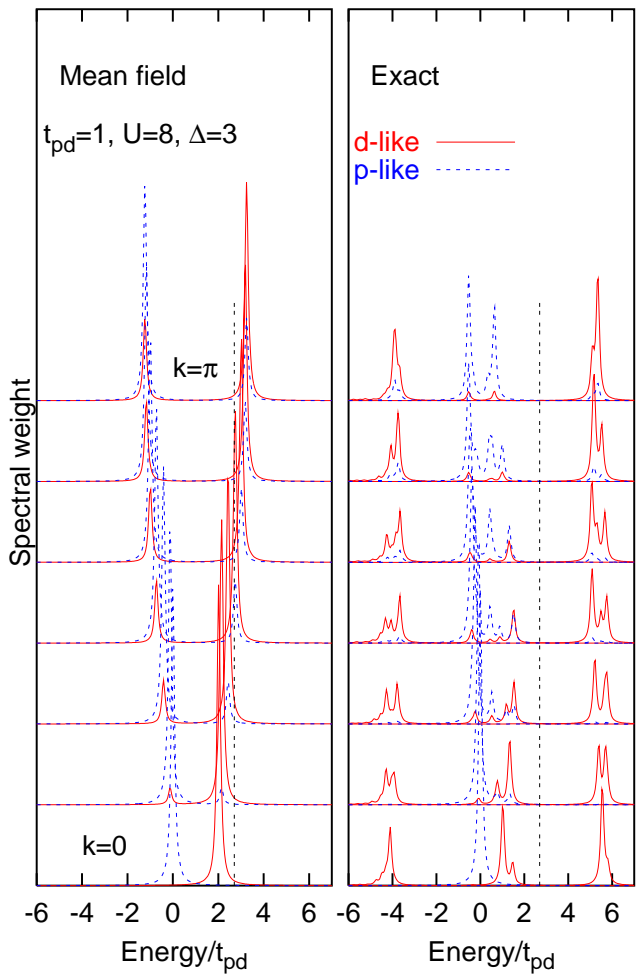}
\caption{\label{fig4} Single particle spectral functions
$A^{(-)}(k,\omega)$ and  $A^{(+)}(k,\omega)$ 
obtained by mean-field solution
of the model and by exact diagonalization of
a system with $6$ unit cells. The wave vector $k$ increases
from the lowermost to the uppermost panel in steps of $\frac{\pi}{6}$.
$\delta$-functions have been replaced by Lorentzians
of width $0.03 t_{pd}$.
The part to the left (right) of the vertical dashed line shows
$A^{(-)}(k,\omega)$ ($A^{(+)}(k,\omega)$).}
\end{figure}

\begin{figure}
\includegraphics[width=\columnwidth]{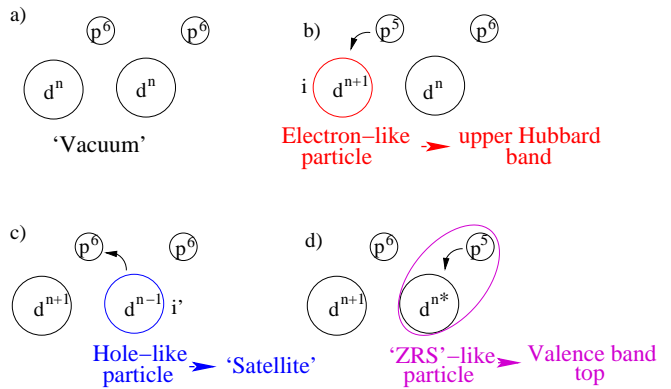}
\caption{\label{fig10} Charge fluctuation processes
relative to the purely ionic configuration.}
\end{figure}

For a start we consider the following minimal version of a 1D
charge transfer model (see Figure \ref{fig3}):
\begin{eqnarray}
H &=& -\Delta \sum_{i,\sigma} d_{i,\sigma}^\dagger d_{i,\sigma}^{}
+ U \sum_{i } d_{i,\uparrow}^\dagger d_{i,\uparrow}^{}
d_{i,\downarrow}^\dagger d_{i,\downarrow}^{}
\nonumber \\
&-&t_{pd} \sum_{i,\sigma}
\left(\;d_{i,\sigma}^\dagger (\;p_{i+\frac{1}{2},\sigma}^{} 
- p_{i-\frac{1}{2},\sigma}^{}\;) + H.c.\;\right)
\nonumber \\
&+&t_{pp} \sum_{j,\sigma}
\left(\;p_{j,\sigma}^\dagger p_{j+1,\sigma}^{} 
 + H.c.\;\right)
\label{model}
\end{eqnarray}
This model describes a 1D chain consisting of strongly correlated
metal($d$) orbitals and uncorrelated ligand ($p$) orbitals.
Henceforth we choose $t_{pd}$ as the unit of energy
and unless otherwise stated set $t_{pp}=0$.
The relevant filling (which we will consider henceforth)
of the model is $3$ electrons (or $1$ hole) per unit cell.
While our goal ultimately is to study realistic models for
compounds such as NiO, our motivation for studying this 
highly oversimplified model is as follows: it is simple enough
so that reasonably large clusters (up to $6$ unit cells)
can be treated by exact diagonalization (ED) and the obtained
exact results for the single-particle spectral function
then can serve as a benchmark for the analytical theory.
The very simple nature of the model thereby is highly
desirable, because it results in  a small
number of `bands' so that the comparison with theory
is more significant than e.g. in the case of NiO.\\
The quantity of main
interest is the photoemission and inverse
photoemission spectrum, defined as
\begin{eqnarray}
A_d^{(-)}(\bm{k},\omega)&=& \sum_\alpha \sum_\mu
|\langle \Psi_\mu^{(n-1)} | d_{\alpha,\bm{k},\sigma}^{} | \Psi_0^{(n)}
\rangle|^2  \nonumber \\
&&\;\;\;\;\;\;\;\;\;\;\;\;\; 
\delta\left( \omega + (E_\mu^{(n-1)} - E_0^{(n)})\right)
\nonumber \\
A_d^{(+)}(\bm{k},\omega)&=& \sum_\alpha \sum_\nu
|\langle \Psi_\nu^{(n+1)} | d_{\alpha,\bm{k},\sigma}^{\dagger} | \Psi_0^{(n)}
\rangle|^2 \nonumber \\
&&\;\;\;\;\;\;\;\;\;\;\;\;\;
\delta\left( \omega- (E_\nu^{(n+1)} - E_0^{(n)})\right)
\label{specdef}
\end{eqnarray}
where $\alpha \in \{ xy, xz, yz,\dots\}$ denotes the type of d-orbital,
and $\Psi_\nu^{(n)}$ ($E_\nu^{(n)}$) denote the $\nu^{th}$ eigen state
(eigen energy) with $n$ electrons - thereby $\nu=0$ corresponds
to the ground state. The spectral function for $p$-electrons
is defined in an analogous way.\\
To get an idea how to construct an adequate theory
it is useful to compare
the paramagnetic mean-field solution
of the model 
(i.e. with $\Delta\rightarrow \Delta_{MF}= \Delta+\langle n_\sigma
\rangle U$) and the results of an ED calculation, see Figure
\ref{fig4}.
The mean-field solution gives two bands
\[
E_\pm(k)= \frac{\Delta_{MF}}{2} \pm\sqrt{
\left(\frac{\Delta_{MF}}{2}\right)^2 + 4 t_{pd}\sin^2\left(\frac{k}{2}\right)}.
\]
The lower (fully occupied) one of these has predominant $p$-character, the
upper (half-occupied) one has predominant $d$-character.
In the spectra obtained by exact diagonalization,
the $p$-like bands persists with an almost
unchanged dispersion. This is not really surprising,
because an electron in the respective state moves
predominantly on the $p$-sublattice and thus will not
feel the strong Coulomb interaction on the
$d$-sites very much. 
The band with predominant $d$-character, on the other hand
disappears completely in the exact spectra.
There is a diffuse band somewhat below $-\Delta=-3t_{pd}$ and a second 
one at the energy $-\Delta+U=5t_{pd}$.
Clearly, these two resemble the `Hubbard bands' expected for
a strongly correlated system and the respective
final states have the character of an empty or doubly occupied
$d$-orbital.  In addition to these Hubbard bands, however,
there is a third group of peaks at energies $\approx+t_{pd}$
which energetically is close to the
$p$-like band, has a mixed $p$-$d$ character
and which does in fact form the first ionization
states of the system. Its closeness to the
$p$- band would seem to suggest that the final states
have a hole predominantly on the $p$-sites, but it also has
a significant admixture of $d$-weight and moreover
is closer to the Fermi energy than the $p$-like band. 
In the charge transfer system under consideration,
the $d$-like band thus actually splits up
into {\em three bands}, rather than the two Hubbard bands which
one might expect.\\
Despite the highly oversimplified nature of the
1D model, there is actually already a clear analogy to NiO:
LDA bandstructure calculations\cite{Shen_long}  produce two
well-separated band complexes, the lower one
(i.e.: the one more distant from the Fermi energy) with
predominant oxygen character, the upper one with Ni character.
This is quite similar to the mean-field solution in
Figure \ref{fig4}.
The actual photoemission spectra, however, show first of all
a broad structure at binding energies $>8 eV$ below the
top of the valence band. Resonant photoemission 
experiments\cite{Ohetal,Tjernberg}
show,  that the final states observed in this energy range have
predominantly $d^7$ character - clearly
they should be identified with the $d$-like band at $-4t_{pd}$ in our model.
Next, at binding energies $-6eV$ and $-4eV$ ARPES 
experiments\cite{Shen_long} find
a group of strongly dispersive bands
which closely resemble the oxygen-like bands obtained
from an LDA band structure calculation
- they obviously
correspond to the dispersive $p$-like `remnant' of the
free-electron band in our model.
Finally, the top of the valence band in NiO is formed by
a group of almost dispersionless bands, whereby the mixed Ni $d^7$ and 
$d^8\underline{L}$ character of these states is established
by resonant photoemission\cite{Ohetal,Tjernberg} - 
these states then would correspond 
to the low intensity band which forms the
top of the electron annihilation spectrum. The cluster calculation of 
Fujimori and Minami\cite{FujimoriMinami} suggests, 
that these states should be viewed
as hole-like `compound objects' where a hole on oxygen is bound to
an excited state of $d^8$ - a type of state that might be viewed as
a generalization of the Zhang-Rice singlet in the CuO$_2$ planes
of cuprate superconductors. As already conjectured by
Fujimori and Minami their `compound nature' would make these
quasiholes very heavy, which immediately would explain the
lack of dispersion seen in the ARPES spectra\cite{Shen_long}.
Finally, inverse photoemission shows
the presence of an upper Hubbard band in NiO
which is also present in the spectra of the 1D model.\\
The above comparison shows, that in addition to $p$-like holes
we will need three types of `effective particles'
to reproduce the correlated band structure of the model.
The two standard Hubbard bands, which correspond to
$d^{n\pm 1}$-like final states are not sufficient here.
To get an idea what these states should be, let us start from the 
ionic limit, $t_{pp}=t_{pd}=0$.
The ground state then corresponds to a constant
number of electrons, $n$, in each metal $d$-shell and completely 
filled ligand $p$-shells, see Figure \ref{fig10}a.
Switching on the hybridization integral
$t_{pd}$ then will produce charge fluctuations:
in a first step, a hole is transferred into a $p$ orbital,
thus producing a $d^{n+1}$ state in $d$-orbital
number $i$, see Figure \ref{fig10}b. The $d^{n+1}$ state has an energy of
$U-\Delta$ relative to the original $d^n$ state, and
will become our first `effective particle' - these
`particles'  form the unoccupied Hubbard band.
In a second step,
the $p$-like hole can be transferred into a $d$-orbital $i'\ne i$,
thus producing a $d^{n-1}$ state, see Figure \ref{fig10}c.
The latter has an energy of
$+\Delta$ relative to the $d^n$ state and provides the
second type of `effective particle' - actually the one
that forms the `satellite' in the spectral function.
Finally, the hole in $i'$ can be transferred
back into a neighboring $p$-level, thereby leaving the
orbital $i'$ in a state $d^{n*}$, i.e. an eigenstate
of $d^n$ other than the orginal one, see Figure \ref{fig10}d.
The `compound object' consisting
of $d^{n*}$ and a hole in a neighboring
$p$-orbital will be the third type of effective particle,
its energy relative to the original $d^n$ state is
$\approx 0$, i.e. appropriate to give top of the valence band.
These states maight be viewed
as generalizations of a Zhang-Rice singlet, or an extreme
case of an (either spin- or orbital-like) `Kondo object'.
Level repulsion due to hybridization
between the $d^{n-1}$ and the $d^{n*}\underline{L}$ states will push 
the latter up to higher energies relative to
the $p$-like bands - in this way, these states
become the first ionization states.
In the following, we will try to give the above considerations
a more solid theoretical foundation and apply them to
the calculation of `correlated band structures'.
\section{General Theory}
We consider a typical transition metal oxide and restrict
our basis to the oxygen $2p$-orbitals and transition metal
$d$-orbitals. 
Taking the energy of the $p$-level as the zero of energy
the single-particle terms in the Hamiltonian then take the form
\begin{eqnarray}
H_{pp}&=&\sum_{i,\kappa,j,\kappa'} \sum_\sigma\;
(t_{i,\kappa}^{j,\kappa'}\;  p_{i,\kappa,\sigma}^\dagger\;
p_{j,\kappa',\sigma}^{} + H.c.),\nonumber \\
H_{pd}&=&\sum_{i,\alpha,j,\kappa} \sum_\sigma\;
(t_{i,\alpha}^{j,\kappa}\;  d_{i,\alpha,\sigma}^\dagger\;
p_{j,\kappa,\sigma}^{} + H.c.),\nonumber \\
H_{dd}&=&\sum_{i,\alpha,j,\alpha'} \sum_\sigma\;
(t_{i,\alpha}^{j,\alpha'}\;  d_{i,\alpha,\sigma}^\dagger\;
d_{j,\alpha',\sigma}^{} + H.c.),\nonumber \\
H_{d}&=&  \sum_{i,\alpha,\beta}\sum_\sigma\;
(V_{\alpha,\beta}\; d_{i,\alpha,\sigma}^\dagger\; d_{i,\beta,\sigma}^{}
+ H.c.),
\end{eqnarray}
where $d_{i,\alpha,\sigma}^\dagger$ creates a spin-$\sigma$ electron in the
$d$-orbital $\alpha\in \{ xy, xz, yz,\dots\}$ on metal site $i$
and $p_{j,\kappa,\sigma}^\dagger$ creates an electron
in $p$-orbital $\kappa\in \{ x,y,z\}$ on oxygen site $j$.
The $V_{\alpha,\beta}$ combine charge-transfer energy and
crystalline electric field.
The Coulomb interaction between the $d$-electrons is
\begin{equation}
H_{C}=\sum_{\zeta_1,\zeta_2,\zeta_3,\zeta_4}
V_{\zeta_1,\zeta_2}^{\zeta_3,\zeta_4}\;
 d_{\zeta_1}^\dagger  d_{\zeta_2}^\dagger
d_{\zeta_3}^{} d_{\zeta_4}^{},
 \end{equation}
where we have suppressed the site label $i$,  $\zeta=(m,\sigma)$
and $m \in \{-2...2\}$ denotes the
$z$-component of the orbital angular momentum. The
matrix-elements
$V_{\zeta_1,\zeta_2}^{\zeta_3,\zeta_4}$
can be expressed in terms of
the $3$ Racah-parameters, $A$ $B$ and $C$.\\
For a start we take all the hybridization matrix elements $t$
to be zero. In this limit
each oxygen is in a $2p^6$ configuration, and
each transition metal ion in one of the ground states
of $H_{intra}=H_d+H_{C}$ with $n$ electrons.
In general, this ground state is degenerate, and
we deal with this by choosing one of these ground states,
which we call $|\Phi_{i,0}\rangle$, for each metal ion -
for example in the case of $NiO$
we would choose the direction of the spin $S=1$ of the
$d$-shell to oscillate between the two sublattices, so as to
describe the antiferromagnetic order in the
system. We call $|\Phi_{i,0}\rangle$ the corresponding
`reference state' on transition metal site $i$,
it obeys $H_{intra} |\Phi_{i,0}\rangle = E_0^{(n)} |\Phi_{i,0}\rangle$.
In the following, we consider the product state
of the $|\Phi_{i,0}\rangle$ and the completely filled
oxygen-$p$ sublattice as the `vacuum' of our theory.
For the 1D model (\ref{model}) we introduce two
$d$-sublattices $A$ and $B$ and choose
\begin{eqnarray*}
|\Phi_{i,0}\rangle&=&d_{i,\uparrow}^\dagger,    \;\;\;\;\;\;\;\;\;\;i\in\;A\\
|\Phi_{i,0}\rangle&=&d_{i,\downarrow}^\dagger,  \;\;\;\;\;\;\;\;\;\;i\in\;B
\end{eqnarray*}
so as to model the antiferromagnetic spin correlations in the
system.\\
Next, we assume that the hybridization between the sub-systems is
switched on. This will create charge fluctuations
in the vacuum: in a first step
an electron from a $p$-shell will be transferred into
one of the $d$-orbitals of the neighboring metal ion $i$, 
a process frequently denoted as $d^n \rightarrow d^{n+1} \underline{L}$.
Due to the many-body character of the Hamiltonian,
the resulting state, $d_{i,\alpha,\sigma}^\dagger |\Phi_{i,0}\rangle$, in general
is not an eigenstate of $H_{intra}$ - rather, we can
express it as a superposition of eigenstates:
\begin{equation}
d_{i,\alpha,\sigma}^\dagger |\Phi_{i,0}\rangle = \sum_\nu\; C_{i,\alpha,\sigma,\nu}
|\nu\rangle,
\end{equation}
where $|\nu\rangle$, $\nu=0,1,\dots\nu_{max}$
are the eigenstates of the $d$-shell with
$n+1$ electrons. Since we are dealing with a single $d$-shell
these can be obtained by exact diagonalization, which
gives us the eigenenergies $E_\nu^{(n+1)}$
of the states $|\nu\rangle$ as well as the coefficients
$C_{i,\alpha,\sigma,\nu}$. If the $|\nu\rangle$ are
chosen to be eigenstates of $S_z$ (as we will assume in all
that follows)
only a small fraction of the $ C_{i,\alpha,\sigma,\nu}$ is
different from zero.
We now represent the state where the metal ion $i$
is in the state $|\nu\rangle$ by the presence of a
`book keeping Fermion', created by $e_{i,\nu,\sigma}^\dagger$, at the site
$i$. The spin index $\sigma$ thereby gives the
difference in $z$-Spin between the state $|\nu\rangle$ and the
reference state
$|\Phi_{i,0}\rangle$ - in principle this is redundant, but we add it so as 
to make the analogy with a free-particle Hamiltonian more obvious.
An important technical point is,
that in case the $|\nu\rangle$ are not eigenstates of $S_z$
this labelling is not possible - there may exist states
$|\nu\rangle$ which can be reached by
transferring an electron of either spin direction
into the $d$-shell $i$.\\
All in all, the charge fluctuation process then can be described by
the Hamiltonian
\begin{eqnarray}
H_1&=& \sum_{i,\nu,\sigma} \epsilon_{i,\nu}\; e_{i,\nu,\sigma}^\dagger\;
e_{i,\nu,\sigma}^{} \nonumber \\
&& \;\;\;+  
\sum_{i,\nu} \sum_{j,\kappa,\sigma} (V_{j,\kappa,\sigma}^{i,\nu}\;
e_{i,\nu,\sigma}^\dagger\; p_{j,\kappa,\sigma}^{}  + H.c.),\nonumber \\
\epsilon_{i,\nu}&=&E_\nu^{(n+1)} - E_0^{(n)}, \nonumber \\
V_{j,\kappa,\sigma}^{i,\nu}&=&\sum_{\alpha}\; t_{i,\alpha}^{j,\kappa}\;
\langle \nu| d_{i,\alpha,\sigma}^\dagger |\Phi_{i,0}\rangle  \nonumber \\
&=&\sum_{\alpha} t_{i,\alpha}^{j,\kappa}\; C_{i,\alpha,\sigma,\nu}.
\label{H1}
\end{eqnarray}
It is understood that states $|\nu\rangle$ which
have $C_{i,\alpha,\sigma,\nu}=0$ for both directions of $\sigma$
should be omitted from this Hamiltonian.
One can see, that the `bare'
hopping integrals $t_{i,\alpha}^{j,\kappa}$ are multiplied
by the coefficients $C_{i,\alpha,\sigma,\nu}$, which have a modulus
$<1$. The effective Fermions
$e_{i,\nu}^\dagger$ thus in general have a weaker
hybridization with the $p$-Orbitals than the original electrons,
which will naturally lead to some kind of `correlation narrowing`
of all bands of appreciable $d$-character.\\
In the 1D model (\ref{model}) there is only one state with
$n+1$ electrons, namely the state 
$d_{i\uparrow}^\dagger d_{i\downarrow}^\dagger |0\rangle$
This has an energy of $U-2\Delta$, whence $\epsilon_{i}=U-\Delta$.
For a site $i$ on the $A$ sublattice we thus identify
$e_{i\downarrow}^\dagger |vac\rangle = d_{i\downarrow}^\dagger 
d_{i\uparrow}^\dagger |0\rangle$ and the part $H_1$ becomes
\begin{eqnarray}
H_1&=& (U-\Delta) \sum_{i\in A} \; 
e_{i\downarrow}^\dagger\; e_{i\downarrow}^{} 
\nonumber \\
&& - t_{pd} \sum_{i\in A} 
\left(\; e_{i\downarrow}^\dagger\; (\;p_{i+\frac{1}{2},\downarrow}
-p_{i-\frac{1}{2},\downarrow}) +  H.c.\; \right)
\end{eqnarray}
plus an analogous term which describes the charge fluctuations
on the sites of the $B$-sublattice.\\
We proceed to the next type of state which
is admixed by the hybridization.
First, an electron from the metal-ion with $n+1$ electrons
may be transferred back to the oxygen atom with the hole,
i.e. $ d^{n+1} \underline{L} \rightarrow d^n$.
If the metal ion $i$ thereby returns to the reference state $|\Phi_{i,0}\rangle$
this process is described by the term `$H.c.$' in (\ref{H1}).
If the metal ion returns to an $n$-electron state $|\lambda\rangle$
other than $|\Phi_{i,0}\rangle$ we should model this by a Bosonic excitation
$b_{i,\lambda}^\dagger$. Here we will
neglect these latter processes - this is presumably the
strongest and least justified approximation in the present theory.
It implies for example that we are neglecting (in the case of
NiO) the coupling to $d-d$ excitons and
the influence of the quantum spin
fluctuations (spin waves) in the 3D S=1 Heisenberg antiferromagnet
formed by the Ni-moments.\\
A second type of state can be generated by filling the hole
in the oxygen shell $j$ with 
an electron from a $d$-shell $i'\ne i$, that means
$d^{n} \underline{L}\rightarrow d^{n-1}$.
This leaves the $d$-shell on $i'$ in an 
eigenstate $|\mu\rangle $ of $n-1$ electrons,
the net effect is the transfer of an
electron between the $d$-shells $i' \rightarrow i'$, i.e. precisely the
process considered originally by Hubbard. We write
\begin{equation}
d_{i,\alpha,\sigma}^{}  |\Phi_{i,0}\rangle = \sum_\mu\; \tilde{C}_{i,\alpha,\sigma,\mu}
|\mu\rangle
\end{equation}
and model the shell $i'$ being in the $\mu^{th}$
ionization state by the presence
of a hole-like book-keeping Fermion, created by $h_{i',\mu,\sigma}^\dagger$.
In an analogous way as above we arrive at the
following effective Hamiltonian to describe the second type of
charge fluctuation:
\begin{eqnarray}
H_2&=& \sum_{i,\mu,\sigma} \tilde{\epsilon}_{i,\mu}\; 
h_{i,\mu,\sigma}^\dagger\;
h_{i,\mu,\sigma}^{} \nonumber \\
 && \;\;\;\;+\sum_{i,\mu} \sum_{j,\kappa,\sigma} 
(\tilde{V}_{j,\kappa,\sigma}^{i,\mu}
\; p_{j,\kappa,\sigma}^\dagger\;h_{i,\mu,\bar{\sigma}}^\dagger 
+ H.c.),\nonumber \\
\tilde{\epsilon}_{i,\mu}&=&E_\mu^{(n-1)} - E_0^{(n)}, \nonumber \\
\tilde{V}_{j,\kappa,\sigma}^{i,\mu}&=&\sum_{\alpha}\; 
(t_{i,\alpha}^{j,\kappa})^*
\;\langle \mu| d_{i,\alpha,\sigma} |\Phi_{i,0}\rangle  \nonumber \\
&=&\sum_{\alpha} (t_{i,\alpha}^{j,\kappa})^*\; \tilde{C}_{i,\alpha,\sigma,\mu}.
\label{H2}
\end{eqnarray}
Since $h_{i,\mu}^\dagger$ creates a hole-like particle,
the presence of terms like $h^\dagger p^\dagger$ is
not unusual - these terms describe particle-hole
correlations, not particle-particle correlations as
in BCS theory.\\
In the $1D$ model, the only state with $n-1$ electrons is the
empty site $|0\rangle$, which has the energy $E=0$, whence
$\tilde{\epsilon}=\Delta$. We identify, for a site $i$ on the
$B$ sublattice: $h_{i\uparrow}^\dagger|vac\rangle=|0\rangle$ and
the term $H_2$ reads
\begin{eqnarray}
H_2&=& \Delta \sum_{i \in B}  h_{i\uparrow}^\dagger h_{i\uparrow}^{}
\nonumber \\
&&-t_{pd} \sum_{i \in B} \left(\;
(\;p_{i+\frac{1}{2},\downarrow}^\dagger
-p_{i-\frac{1}{2},\downarrow}^\dagger) h_{i\uparrow}^\dagger +
H.c. \right)
\end{eqnarray}
plus an analogous term describing the $A$ sublattice.\\
As already stated, the second type of charge transfer excitation
describes the transfer of an electron between two $d$-shells
via an intermediate state with a hole in an oxygen $p$-shell.
If there are direct $d$-$d$ transfer integrals, this
process can also occur in one step. The respective part of the
effective Hamiltonian can be constructed in an entirely analogous
fashion as the parts above - since it is lengthy, we do
not write it down in detail.\\
We proceed to the last type of `effective particle'
that we will consider.
If the $d$-shell on atom $i$ is
in an eigen state of $n-1$ electrons (which would be described by the 
$h_{i,\mu}^\dagger$-particle) it is possible that an electron from
a neighboring $p$-shell is transferred to the $d$-shell,
thereby leaving the $d$-shell $i$ in an eigenstate $|\lambda\rangle$
of $n$ electrons other than the reference state $|\Phi_{i,0}\rangle$.
We will consider the `compound object'
consisting of the `excited' $d^n$ state $|\lambda \rangle$ on site
$i$ and a hole in a linear combination $\psi_{\alpha,\sigma}^\dagger$
of $p$-orbitals on the nearest neighbors of $i$
as a further effective particle, created by
 $z_{i,\lambda,\alpha,\sigma}^\dagger$.
Here $\alpha \in \{ xy, xz, yz,\dots\}$ denotes the
symmetry of the linear combination of
$p$-orbitals, which is such as to hybridize
with exactly one of the $d_{i,\alpha,\sigma}^\dagger$ on
site $i$. For an ideal tetrahedral cage of oxygen
atoms around each metal ion there is exactly one combination
$\psi_{\alpha,\sigma}^\dagger$ for each $\alpha$.
The creation and annihilation
of the $z$-particles can be described by the term
\begin{eqnarray}
H_3&=& \sum_{i,\lambda,\alpha,\sigma} \epsilon_{i,\lambda,\alpha}\; 
z_{i,\lambda,\alpha,\sigma}^\dagger\; z_{i,\lambda,\alpha,\sigma}^{} \nonumber \\
&& \;\;\;+  \sum_{i,\mu,\lambda,\alpha}
(V_{i,\lambda,\alpha,\sigma}^\mu \;
z_{i,\lambda,\alpha,\sigma}^{\dagger} h_{i,\mu,\sigma} + H.c.),\nonumber \\
\epsilon_{i,\lambda,\alpha}&=&E_\lambda^{(n)} - E_0^{(n)} -\zeta_\alpha, 
\nonumber \\
V_{i,\lambda,\alpha,\sigma}^{\mu}&=& - T_{\alpha}\;
\langle \lambda| d_{\alpha,\sigma}^\dagger |\mu \rangle  
\label{H3}
\end{eqnarray}
Here $\zeta_\alpha= \langle 0|
\psi_{\alpha,\sigma}[H_{pp},\psi_{\alpha,\sigma}^\dagger]|0\rangle$ 
denotes the kinetic energy of the combination
$\psi_{\alpha,\sigma}^\dagger$ - it can be expressed in terms of the
integrals $(pp\sigma)$ and  $(pp\pi)$.
Also, $T_{\alpha}= \langle 0| d_{i,\alpha,\sigma}
[ H_{pd}, \psi_{i,\alpha,\sigma}^\dagger] | 0 \rangle$
is the hybridization matrix element
between $\psi_{\alpha,\sigma}^\dagger$ and
$d_{i,\alpha,\sigma}^\dagger$ and can be written in terms of
$(pd\sigma)$ and  $(pd\pi)$.\\
In the 1D model, the only `excited` state on
the $A$ sublattice is the state $d_{i,\downarrow}^\dagger |0\rangle$
The only possible combination of $p$ orbitals which
hybridizes with a $d$-orbital is
$\psi_{i,1,\sigma}^\dagger
=\frac{1}{\sqrt{2}}((\;p_{i+\frac{1}{2},\sigma}^{\dagger}  
-p_{i-\frac{1}{2},\sigma}^{\dagger}\;)$, which has
$\zeta_1=-t_{pp}$. 
There is therefore just one
$z^\dagger$-like particle on the $A$-sublattice,
which corresponds to site $i$ being  in the state
$d_{i,\downarrow}^\dagger |0\rangle$ and having an extra
hole in the combination $\psi_{i,1,\sigma}^\dagger$.
The total energy is
$E=-\Delta+t_{pp}$ whence $\epsilon=t_{pp}$ and
the corresponding Hamiltonian reads
\begin{equation}
H_3 = t_{pp} \sum_{i\in A} z_{i\uparrow}^\dagger z_{i\uparrow}^{}
- \sqrt{2}t_{pd} \sum_{i\in A} (z_{i\uparrow}^\dagger h_{i\uparrow}^{} + H.c.)
\label{H3_simp}
\end{equation}
and, again, a corresponding term for the $B$-sublattice.
This concludes the types of state which we take into account.
We are thus assuming that the hole always is
on a nearest neighbor of the $d^{n*}$ state - this means that
we truncate the `Kondo cloud' which is not
exactly true. In principle this approximation could be
relaxed by including more complex composite particles
but here we do not include these.\\
In order for the mapping between the actual system and
the `book-keeping Fermions' to be a faithful one, we must require
that the occupation of any $d$-site is either $0$ or $1$ -
otherwise, the state of the respective $d$-shell is not unique.
This implies that the book-keeping Fermions $e^\dagger$,
 $h^\dagger$ and  $z^\dagger$ have to obey a hard-core
constraint, in exactly the same way as e.g. the
magnons in spin-wave theory for the Heisenberg antiferromagnet (HAF).
It is well known, however, that linear spin-wave theory for the HAF, which
neglects this hard-core constraint alltogether and treats the magnons as free
Bosons, gives an excellent description of the antiferromagnetic
phase, even in the case of
$S=1/2$ and $d=2$, where quantum fluctuations are strong. 
The reason is, that the density $n$ of magnons/site obtained 
self-consistently from linear spin wave theory is still
relatively small, whence the probability that two magnons
occupy the same site and thus violate the constraint is
$\propto n^2 \ll 1$. For the same reason we expect that
relaxing the constraint in the present case and treating
the book-keeping Fermions as free Fermions will be a very good
approximation - its physical content is the assumption, that
the probability of charge fluctuations is small, which is
certainly justified in a Mott- or charge-transfer-insulator.
For completeness we note that there is also
a certain interference between the
$z^\dagger$ particle and the holes on oxygen in the sense that
the respective creation and annihilation operators do not
exactly anticommute. Again, we neglect this, with the
justification again being the very low density of the
$z^\dagger$ and $p^\dagger$ particles.\\
Adding the various terms in the Hamiltonian and the
direct $p-p$ hopping $H_{pp}$
we obtain a Hamiltonian which describes the lowest order charge fluctuation
processes while still being readily solvable by Fourier
and Bogoliubov transform with the result:
\begin{equation}
H= \sum_{\bf{k},\zeta,\sigma} E_{\bf{k},\zeta,\sigma}
\gamma_{\bf{k},\zeta,\sigma}^\dagger \gamma_{\bf{k},\zeta,\sigma}^{},
\end{equation}
where $\zeta$ is a band index.
Quantities of physical interest now can be readily calculated.
Let us first discuss the electron count.
We assume that the reference states for the $d$-shell have
$n$ electrons each. Then, the total number of electrons/unit cell
is $n_e= n\cdot n_d + 6\cdot n_o$ where
$n_d$($n_o$) denote the number of metal (oxygen) atoms in the unit
cell. On the other hand we have
\begin{eqnarray}
n_e &=&  n\cdot n_d + \frac{1}{N} \sum_{\bm{k},\sigma}\;(\;
\sum_{j,\kappa} p_{\bm{k},j,\kappa,\sigma}^\dagger 
p_{\bm{k},j,\kappa,\sigma}^{}
\nonumber \\
&& \;\;\;\;\;\;
+ \sum_{i,\mu} e_{\bm{k},i,\mu,\sigma}^\dagger e_{\bm{k},i,\mu,\sigma}^{}
-  \sum_{i,\nu} h_{\bm{k},i,\nu,\sigma}^\dagger
h_{\bm{k},i,\nu,\sigma}^{}
\nonumber \\
&& \;\;\;\;\;\;\;\;\;\;\;
-  \sum_{i,\alpha,\lambda} 
z_{\bm{k},i,\alpha,\lambda,\sigma}^\dagger z_{\bm{k},i,\alpha,\lambda,\sigma}^{}\;).
\end{eqnarray}
Here the sums over $j$ and $i$ run over the $p$ and $d$-shells
in one unit cell and the equation follows readily
from the electron/hole-like character of the
various effective Fermions. 
This can be rewritten as
\begin{eqnarray}
n_e
&=&  n\cdot n_d + \frac{1}{N} \sum_{\bm{k},\zeta,\sigma}
\gamma_{\bm{k},\zeta,\sigma}^\dagger \gamma_{\bm{k},\zeta,\sigma}^{}
\nonumber \\
&&\;\;\;\;\;\;\;\;\;\;\; - 
\sum_{\sigma} (\mu_{tot,\sigma}+\lambda_{tot,\sigma}).
\end{eqnarray}
where $\mu_{tot,\sigma}$ denotes the total number of
ionization states in the unit cell which can be reached by
extracting a spin $\sigma$ electron from one of the
$|\Phi_{i,0}\rangle$, and 
and $\lambda_{tot,\sigma}$ the total number of $z^\dagger$-particles
in the unit cell which couple to one of these
ionization states. If we assume that the band structure is
spin independent (which is the case for NiO) these numbers must be independent
of $\sigma$ and we
obtain the following requirement for the electron number:
\begin{equation}
\sum_{\bm{k},\sigma}
\gamma_{\bm{k},\alpha,\sigma}^\dagger \gamma_{\bm{k},\alpha,\sigma}^{}
= N\left(\; 6 n_o + 
2 (\mu_{tot}+\lambda_{tot})\; \right).
\end{equation}
The number of occupied bands in the system thus is
$n_{occ}=3n_o + \mu_{tot}+\lambda_{tot}$. Since the total
number of bands produced by our formalism
is $3n_o + \nu_{tot}+\mu_{tot}+\lambda_{tot}$, 
we find that the chemical potential
falls exactly into the gap between the
$3n_o + \mu_{tot}+\lambda_{tot}$ bands which correspond 
(in the limit of vanishing
hybridization)  to the oxygen $2p$ states, the 
ionization states of the $d$-shell and the
$d^8\underline{L}$-type states on one hand and the
$\nu_{tot}$ bands, which correspond to the affinity states
of the $d$-shells on the other hand. Obviously, this is the physically
correct position of the chemical potential.\\
The quantity of main interest to us, the single-particle
spectral function $A_{\alpha}(\bm{k},\omega)$
can be obtained from the eigenvectors of the Hamilton matrix
once the resolution of the
$d$-electron creation/annihilation operator is known.
Were it not for the presence of the $z^\dagger$-like particles, we
could write
\begin{eqnarray}
d_{\bf{k},\alpha,\sigma}^\dagger &=& 
\sum_\nu  C_{i,\alpha,\sigma,\mu} \;e_{\bf{k},\mu,\sigma}^\dagger
+ \sum_\nu  \tilde{C}_{i,\alpha,\sigma,\nu}^*\;
h_{-\bf{k},\nu,\bar{\sigma}},
\nonumber \\
\label{resolution}
\end{eqnarray}
which is easily verified by taking matrix-elements
between the right an left-hand side. However, the
presence of the $z^\dagger$-like
`particle' complicates this. Due to their `compound nature'
the processes by which the electron annihilation operator
couples to a $z^\dagger$-particle are rather complicated.
For example, one might envisage a process in which an electron in
a $d^{n+1}$ configuration on site $i$ is annihilated, leaving the 
$d$-shell in a $d^{n*}$ state (i.e. an eigenstate of $d^n$ other than
the reference state $|\Phi_{i,0}\rangle$ on site $i$). Then, if
simultaneously a hole happens to
be present in a $p$-orbital next to site $i$,
this process would create a $z^\dagger$-like particle on
site $i$, leading to an operator product of the type
$z_i^\dagger h_i \psi_{i\alpha}$ to describe this process.
Similarly, if a  $d^{n*}$ state is somehow created
on a site $i$ (this is not possible in the framework of the
Hamiltonian which we wrote down above - it would necessitate
terms including the Bosonic excitations $b_{i,\lambda}^\dagger$ 
discussed above)
and a hole is created in a $p$-orbital next to this site,
this would result in the creation of a $z^\dagger$-like particle on
site $i$. This process could be described by a product
of the type $z_i^\dagger \Psi_{i\alpha} b_i$.
Due to the fact that these processes all involve products of three
operators one might expect that they
lead predominantly to an incoherent continuum in the spectral
function. All in all,
we may thus expect that this type of process will not
contribute substantially to the photoemission intensity
for the dominant peaks. Clearly, the problem in calculating the
spectral weight is a drawback of the theory - it should be noted,
however, that there is a very clear physical reason for this problem,
namely the `compound nature' of the $z_i^\dagger$ particles
and this should be reflected in any theoretical description.
\begin{figure}
\includegraphics[width=\columnwidth]{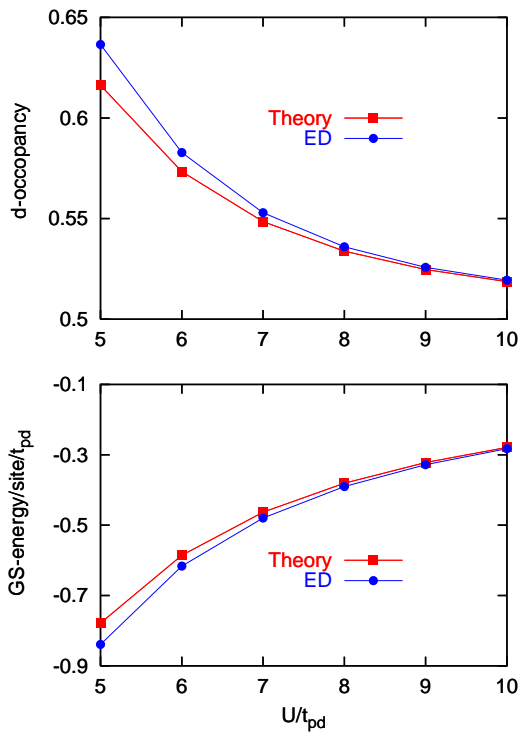}
\caption{\label{fig9} Ground state energy and $d$-occupation
as a function of $U$ for $\Delta=3$, $t_{pd}=1$ as obtained by
exact diagonalization of a system with $6$ unit cells and from the
present theory. The trivial contribution of $-\Delta$ has been
subtracted off from the ground state energy.}
\end{figure}
\section{Comparison with exact diagonalization}
Our theory involves a number of strong approximations
which need to be checked in some way. Here we present a
comparison of results obtained for the 1D model (\ref{model})
and exact diagonalization of finite clusters.
For the 1D model systems with $6$ unit cells easily can be
solved exactly on a computer and we use these results as
a benchmark to check our theory. The simplicity of the
model actually makes the comparison more significant than e.g. in the
case of a realistic model for NiO, because we expect only a small
number of `bands' whence any disagreement will be more obvious.\\
Before we discuss results for physical quantities
let us address one of the key approximations of our theory, namely the
neglect of the hard-core constraint which in principle
should be obeyed by the book-keeping Fermions. Solving the
1D model with $\Delta=3$, $U=6$ gives the GS expectation values
$\langle d_i^\dagger d_i^{} \rangle=0.151$,
$\langle h_i^\dagger h_i^{} \rangle=0.004$ and
$\langle z_i^\dagger z_i^{} \rangle=0.0007$. This implies that the
probability for a violation of the constraint on any given
$d$-site is $\approx 0.03$, i.e. entirely negligible.
Simply relaxing the constraint thus is probably an excellent
approximation.\\
Next, Figure \ref{fig9} compares the total energy/site
and the $d$-occupancy, which is a measure for the charge-transfer
form $p\rightarrow d$, as obtained by exact diagonalization
and from the theory. Obviously, there is good agreement.
\begin{figure}
\includegraphics[width=\columnwidth]{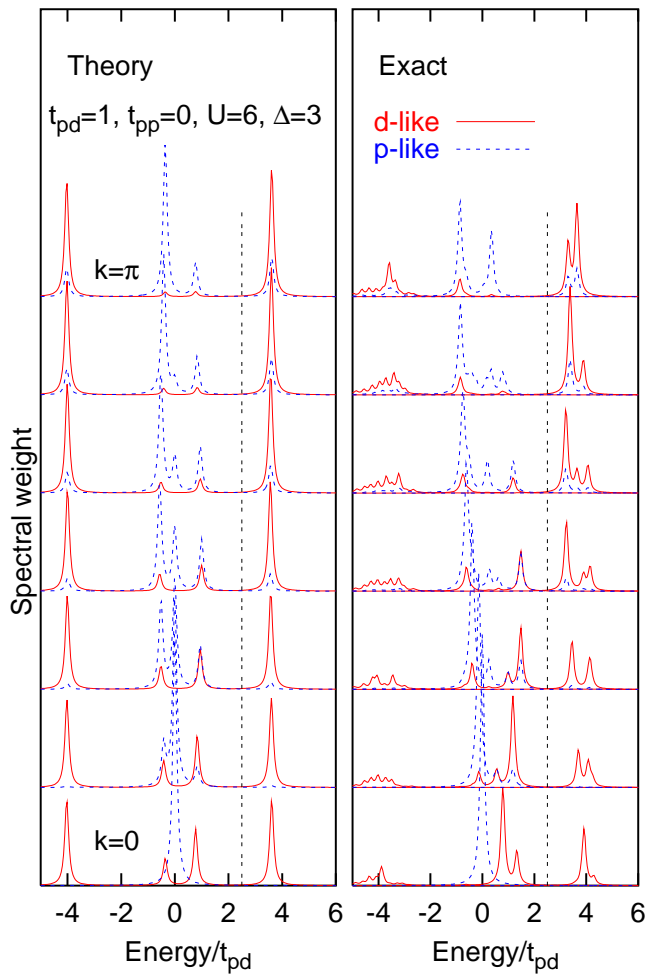}
\caption{\label{fig5} Single particle spectral functions
$A^{(-)}(k,\omega)$ and  $A^{(+)}(k,\omega)$ 
obtained by the present theory and by exact diagonalization of
a system with $6$ unit cells. The wave vector $k$ increases
from the lowermost to the uppermost panel in steps of $\frac{\pi}{6}$,
to that end the figure combines spectra
obtained with periodic and antiperiodic boundary conditions.
The part to the left (right) of the vertical dashed line shows
$A^{(-)}(k,\omega)$ ($A^{(+)}(k,\omega)$).}
\end{figure}
Figure \ref{fig5} shows a comparison between the single particle
spectral function obtained from the theory and by exact
diagonalization of a system with $6$ unit cells. To
obtain a denser mesh of $k$-points, 
spectra for a system with periodic and antiperiodic boundary
conditions have been used for the exact diagonalization
part, that means $k=0,\frac{\pi}{3},\frac{2\pi}{3}$ and $\pi$
have been calculated with PBC, the ones for
 $k=\frac{\pi}{6},\frac{\pi}{2}$ and $\frac{5\pi}{6}$
have been obtained with ABC. Although there is no rigorous proof
for this, experience shows that combining spectra with PBC and ABC
gives quite `smooth' dispersion relations - as can also be seen in
the present case.
In order to suppress the Luttinger-liquid behaviour expected
for $1D$ systems, a staggered magnetic field of $0.1 t_{pd}$
was applied.
The agreement between theory and exact diagonalization
then is obviously quite good. The dispersion and spectral character
of the main `bands' in the numerical spectra is reproduced quite well.
The main difference concerns the very strong damping of the lower
Hubbard band at $E\approx-4t_{pd}$,
which actually forms broad continuum rather than a
well-defined band in the numerical spectra.
Moreover, the upper Hubbard band at $E\approx 4t_{pd}$ 
has some `fine structure' in the numerical
spectra, which is not reproduced by the theory.
On the other hand, our theory does not include any damping mechanism such as
the coupling to spin excitations, so one cannot expect it
to reproduce such details.
Another slight discrepancy concerns the
bandwidth of the oxygen band at  $E\approx 0$, which
is somewhat underestimated by theory. Apart from that and
a few low-intensity peaks in the numerical spectra, however,
there is a rather obvious one-to-one correspondence
between the bands in the theoretical spectra and the exact ones.
\begin{figure}
\includegraphics[width=\columnwidth]{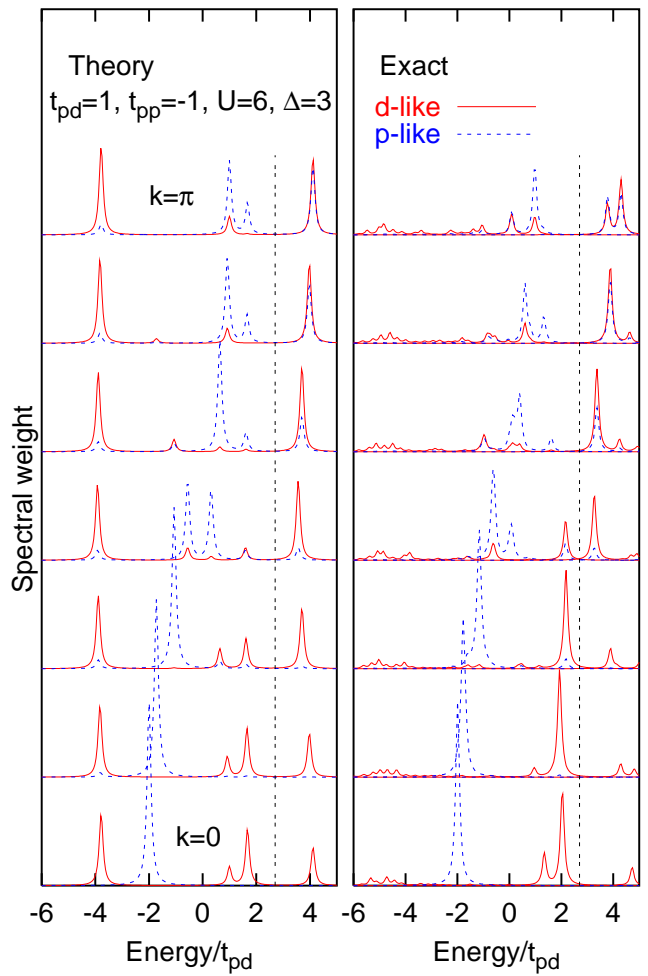}
\caption{\label{fig6} Same as figure (\ref{fig5}) but with different
  parameter
values.}
\end{figure}
Next, we consider the spectra for
a nonvanishing $p$-$p$ hopping, $t_{pp}=-1$, see Figure \ref{fig6}.
Again, there is good agreement between theory and numerics, with the
main discrepancy being again the damping of the satellite and
the fine structure of the upper Hubbard band. 
Still, there is a clear one-to-one correspondence between
theory and exact spectra. An interesting check
is provided by inverting the sign of $t_{pp}$. One might expect
at first sight that the
only effect is to invert the dispersion of the $p$-like band.
Inspection of the Hamiltonian (\ref{H3_simp}) shows, however, that inverting
the sign of $t_{pp}$ also affects the energy
of the $z^\dagger$-particle, and
hence should lead to a shift of the corresponding
band. The actual
spectra in Figure \ref{fig7} then show, that this is
indeed the case in the numerical spectra. The $z$-like band is shifted
to higher energies by very nearly the amount of
$2t_{pd}$ expected from theory, so that the lowest hole-addition
states now belong to the $p$-like band. The fact that the inversion of
the sign of $t_{pp}$ has precisely the effect predicted by
theory is a strong indication, that this is indeed
the correct interpretation of the low energy peaks in the spectra.
\begin{figure}
\includegraphics[width=\columnwidth]{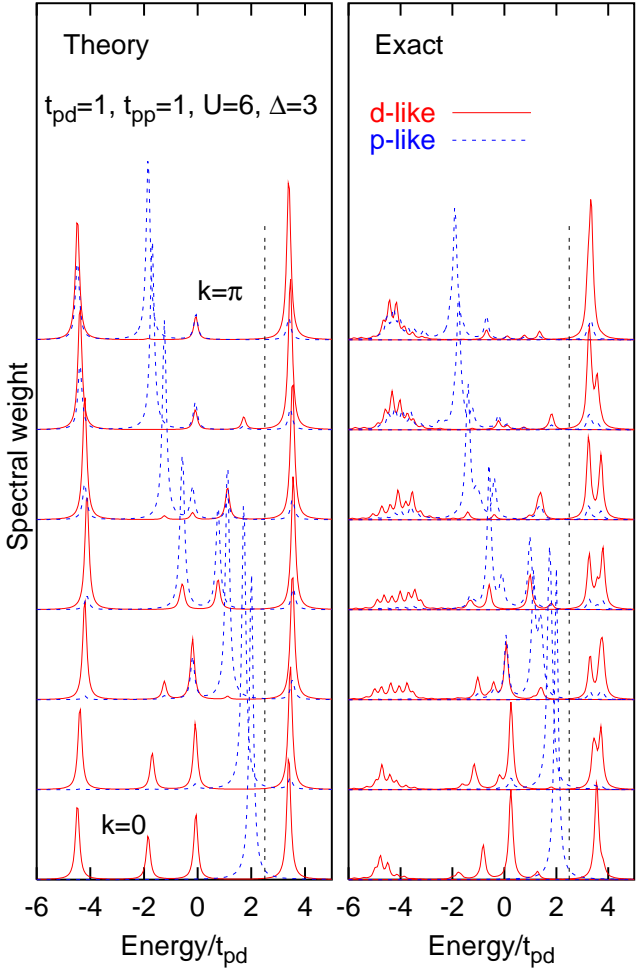}
\caption{\label{fig7} Same as figure (\ref{fig5}) but with different
  parameter
values.}
\end{figure}
\section{The band structure of NiO}
Summarizing the results of the preceeding section we may say that the theory
reproduces the numerical spectra and the trends under a change of
parameters remarkably well, an indication that despite its simplicity 
the theory really captures the essential physics of the two-band model.
This is encouraging to apply it to a real material, NiO.
In applying the above procedure to NiO we first performed a standard 
LDA band structure calculation in the framework of the
LMTO-method\cite{Andersen} for NiO (thereby assuming a paramagnetic
ground state) and obtained the LCAO parameters by a fit. 
For simplicity no overlap integrals were taken into
account. 
\begin{figure}
\includegraphics[width=\columnwidth]{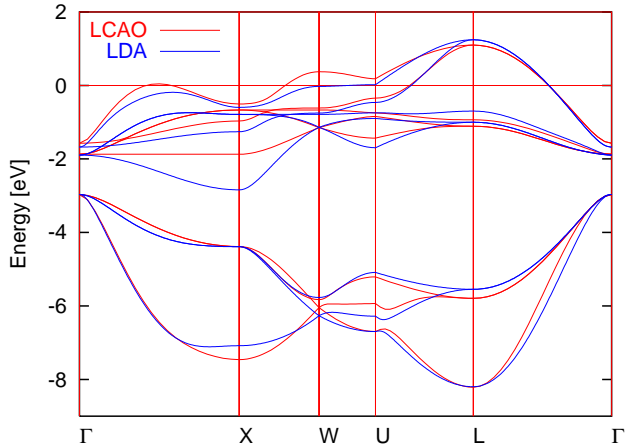}
\caption{\label{fig1} LDA band structure for paramagnetic NiO 
and LCAO fit. The Fermi energy is taken to be zero.}
\end{figure}
A comparison between the LDA band structure and the LCAO fit is shown in
Figure \ref{fig1} (the LDA result is essentially identical to that of
Ref.\cite{Shen_long}), the hybridization integrals and site
energies obtained by the fit are given in Table \ref{tab1}.
We have also obtained
LCAO parameters for an antiferromagnetic LSDA band structure, and those
parameters which can be compared (such as the hybridization
integrals) do not differ significantly.
All in all this procedure gives quite reliable estimates for the
values of the various hopping integrals. 
The LDA bandstructure broadly can be divided into two complexes of
bands: the lower one at energies between $-8eV$ and $-3eV$
has almost pure oxygen $p$-character. In other words, a hole
in this bands would move almost exclusively on the
oxygen sublattice and have only a very small probability to
be on a Ni ion. One may thus expect, that these
states persist essentially unchanged in the the correlated
ground state. Next, the complex between  $-3eV$ and $+1eV$
has almost exclusively Ni $3d$ character. The LDA band structure
thus would seem to suggest, that there are states,
where a hole is moving essentially from one Ni site
to another, which have a less negative binding energy -
i.e. which are closer to the Fermi energy -
than the states where the hole is moving in
the oxygen sublattice. Clearly, in view of the value of
the charge transfer energy $\Delta > 0$, which is
consistently suggested by a variety of
methods\cite{FujimoriMinami,Elp,NormanFreeman}, 
this is a quite wrong picture of the electronic structure. 
\begin{table}[b]
\begin{center}
\begin{tabular}{|c|ccc|l|}
\hline
& Ni-O & O-O & Ni-Ni & $\epsilon$\\
\hline
 $(ss\sigma)$        & -  &    0.023      & - &       \\
 $(sp\sigma)$        & -       & -       & - &  $\epsilon_{2s}=-10$     \\
 $(pp\sigma)$        & -  &    0.665      & - &       \\
 $(pp\pi)$        & -  &   -0.104      & - &    $\epsilon_{2p}=-4.8$   \\
 $(sd\sigma)$   &   -0.720      & -       & - &       \\
 $(pd\sigma)$   &   -1.310      & -       & - &  $\epsilon_{3d}=-1.13$     \\
 $(pd\pi)$   &    0.382      & -       & - &       \\
 $(dd\sigma)$        & -       & -  &   -0.201&       \\
\hline
\end{tabular}
\caption{Hybridization integrals and site-energies (in $eV$)
obtained by a LCAO fit to the paramagnetic LDA band structure of NiO.}
\label{tab1}
\end{center}
\end{table}
Next, we consider the Racah parameters $B$ and $C$. These
differ only slightly from their values for free ions
and we took the values from Fujimori and Minami\cite{FujimoriMinami}
$B= 0.127\;eV$, $C=0.601\;eV$ for $d^8$ and
$B= 0.138\;eV$, $C=0.676\;eV$ for $d^9$. In general, these
parameters are screened by covalency between $d$-orbitals
and ligands\cite{ShulmanSugano} but for simplicity we keep the 
`bare' values.\\
This leaves us with two parameters, which require a special
treatment, namely the Racah parameter $A$, which is subject to
substantial solid-state screening, and the difference of site
energies between the Ni $3d$-level and the
oxygen $2p$-level. 
The Racah parameter $A$ is related to the Coulomb energy $U$, which
can be obtained from `pure $d$-quantities'
according to $U=E_0^{n+1} + E_0^{n-1} -2E_0^n$.
Here we used the values $U=8.7eV$ and $\Delta=1.5eV$.
Similar values for $U$
have been obatined by Fujimori and Minami\cite{FujimoriMinami}
from a cluster fit of the valence band photoemission spectrum,
by van Elp {\em et al.}\cite{Elp} from a cluster fit to the
X-ray absorption spectrum and by
Norman and Freeman\cite{NormanFreeman} from density functional
calculations. The value of $\Delta$ is somewhat small compared
to others, which are around $2.5eV$.\\
\begin{figure}
\includegraphics[width=\columnwidth]{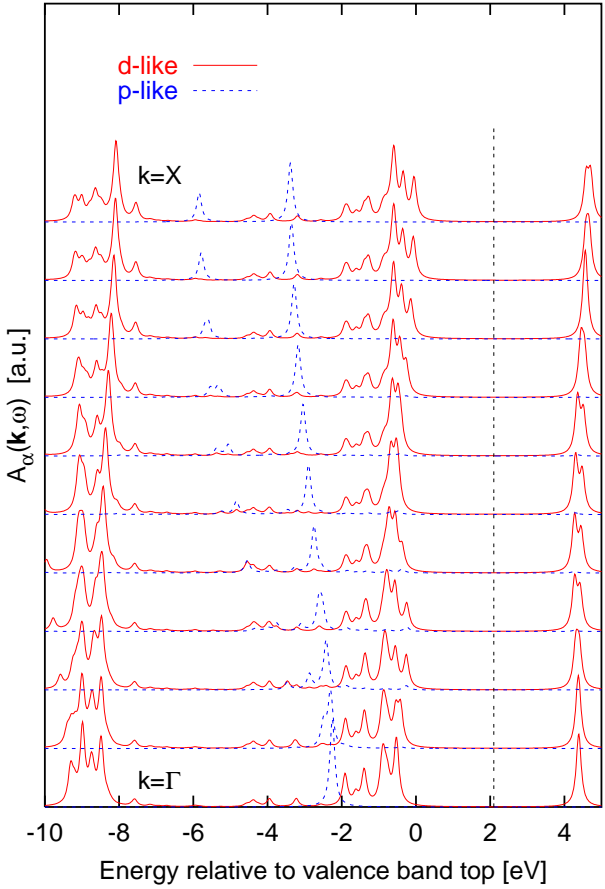}
\caption{ \label{fig11} Single particle spectral densities
(see (\ref{specdef}) for a definition)
for antiferromagentic NiO obtained by the present theory.
The momenta are along the $(100)$ direction,
the top of the valence band is the zero of energy. 
$\delta$-functions are replaced by Lorentzians of width
$0.075eV$, the $d$-like spectral density is multiplied by a factor of $4$.}
\end{figure}
Then, the problem of a single $d$-shell
was solved by exact diagonalization in the $7$, $8$ and $9$-electron
subspaces. The maximum dimension of the
Hilbert space was 120 for $n=7$. 
A nonvanishing CEF parameter $10Dq=0.05eV$ was
applied in order to stabilize the correct $t_{2g}^6 e_g^2$ $^3A_{2g}$
ground state for $d^8$ in $O_h$-symmetry. To account for the antiferromagnetic
nature of the GS of NiO, we chose the reference state $|\Phi_{i,0}\rangle$
to be the $S_z=1$ member of the $^3A_{2g}$ multiplet
on the Ni-sites of one sublattice, and the $S_z=-1$ member on the
other one. Since we neglect spin-orbit coupling the direction
of the spin quantization axis is arbitrary and has no influence
on  the spectral function.
The kinetic energies of the $t_{2g}$ and $e_g$-like combinations
of $p$-orbitals, which enter the energy of the $z$-like particles
are $\epsilon_{t_{2g}}=(pp\sigma)-(pp\pi)$ and
$\epsilon_{e_g}=(pp\pi)-(pp\sigma)$, the respective hybridization
integrals are $T_{t_{2g}}=2(pd\pi)$ and $T_{e_g}=\sqrt{3}(pd\sigma)$.
All in all, the rank of the effective Hamilton matrix to be diagonalized 
was $\approx 250$ i.e. quite moderate.
To improve the agreement with experiment, the following minor
adjustments of parameters were made: the $p-p$ hybridization integrals were
reduced by a factor of $0.8$, and the
$d-p$  hybridization integrals were increased by a factor of
$1.1$.\\
The full single particle spectral functions
obtained along the $(100)$ direction 
for antiferromagnetic NiO then is shown in Figure \ref{fig11}.
\begin{figure}
\includegraphics[width=\columnwidth]{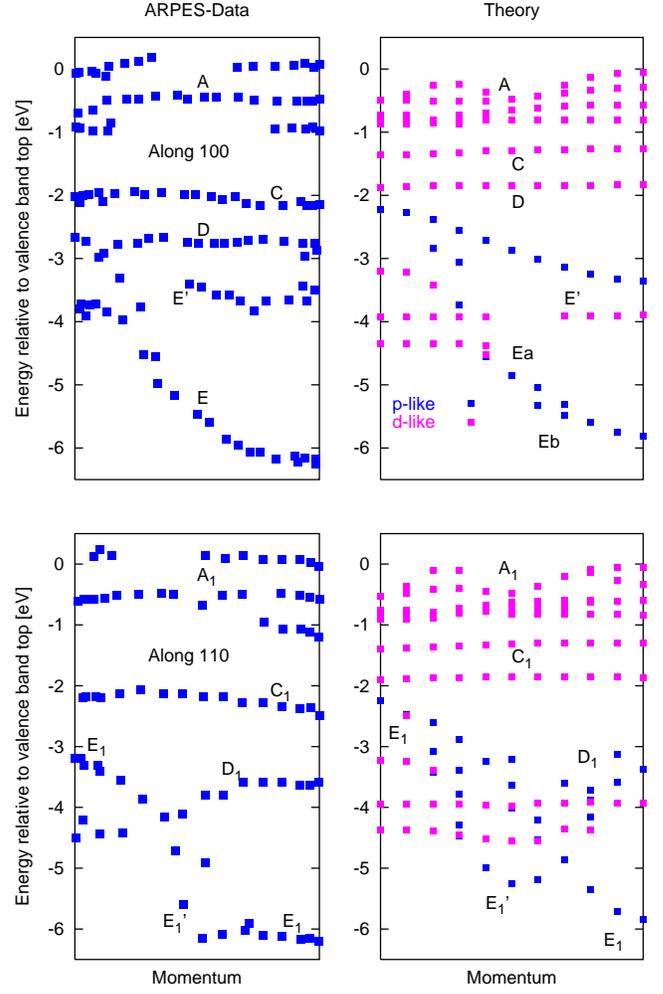}
\caption{\label{fig12} Comparison between the experimental
peak dispersions determined by ARPES in non-normal emission
(taken from Figure 12 of Ref. \cite{Shen_long})
and the position of `significant peaks' in the 
theroretical spectra. The labels on the `bands' indicate
a possible correspondence between experiment and theory.}
\end{figure}
It differs quite significantly from the what one would
expect on the basis of the LDA band structure (see Figure(\ref{fig1}))
but instead shows the same overall structure as in the 1D model,
compare Figure \ref{fig4}.
As was the case for the 1D model one can broadly speaking 
distinguish four complexes of bands.
At binding energies $< -8eV$ , there is a broad continuum
of bands with strong $d$-weight. Analysis of the wave functions
shows, that the respective
states have (mainly) $h^\dagger$ (i.e.: $d^7$) character, with some
admixture of $z^\dagger$ (i.e.: $d^8\underline{L}$) and
(less) admixture of $O2p$ character. Clearly these 
bands should  be identified
with the `satellite' in the experimental NiO spectra.
By analogy with the 1D model
we may expect that these high energy states undergo
substantial broadening as is indeed seen in
experiment. In Figure \ref{fig11} 
the satellite by and large disperses upwards as one goes
away from $\Gamma$ - Shen {\em at al.}\cite{Shen_long}
interpreted their data
as showing a downward dispersion of the satellite.
On the other hand this feature is rather broad and
composed of many `subpeaks' so that it may be difficult to
make really conclusive statements about the dispersion
of the spectral weight without a full
calculation of the spectral weight, including the
`radiation characteristics' of the individual $d$-orbitals,
final states effects etc.\\
Next, there is a group of strongly dispersive
bands of predominant $O2p$ character,
which closely resembles the lower complex of bands in the 
LDA calculation, see Figure \ref{fig1}.
In view of their almost pure oxygen character it is
no surprise that these bands are hardly influenced by whatever happens on
the Ni sites. Next comes a group of practically dispersionless
bands which form the top of the valence band.
They have mainly $z^\dagger$ (i.e.: $d^8\underline{L}$) character with
some admixture of $h^\dagger$ (i.e.: $d^7$).
Due to their strong $z^\dagger$-character these bands probably are
influenced most strongly by our approximation to omit
any terms involving $z^\dagger$-operators in the spectral weight
operator (\ref{resolution}). We may expect that taking the
prcocesses discussed there will probably enhance the weight of
these states and also add some more $p$-like weight to these peaks.\\
The topmost peak is rather intense and actually composed
of several `subpeaks' - it is in fact the only feature in this energy range
which shows significant dispersion. 
Below this broad peak, there are several bands with lower intensity
and practically no dispersion - all of this
exactly as seen in the ARPES experiment by Shen {\em et al.}\cite{Shen_long}. 
Figure  \ref{fig12} shows a more detailed comparison of the
dispersion of `significant peaks' in the photoemission
part of the theoretical spectra with the experimental peak dispersions
as reported by Shen {\em et al.}.
It can be seen, that the agreement is quite good. Along both $(100)$
and $(110)$ the main discrepancy is the position of bands $C$
and $D$ (or $D_1$ along $(110)$) which are
somewhat higher in energy in the theory - still, the discrepancy is
only a fraction of an $eV$. In view of the fact that we
have used the simplemost set of parameters this is quite
good agreement. The band portion $Ea$ which is
unusual due to its downward curvature has actually been
observed by Shen {\em at al.} in normal emission (see Figure 6
of Ref. \cite{Shen_long}). The part $Eb$  seems to correspond
to the experimental band $E$ itself - it has rather low
spectral weight for momenta close to $\Gamma$.\\
\begin{figure}
\includegraphics[width=\columnwidth]{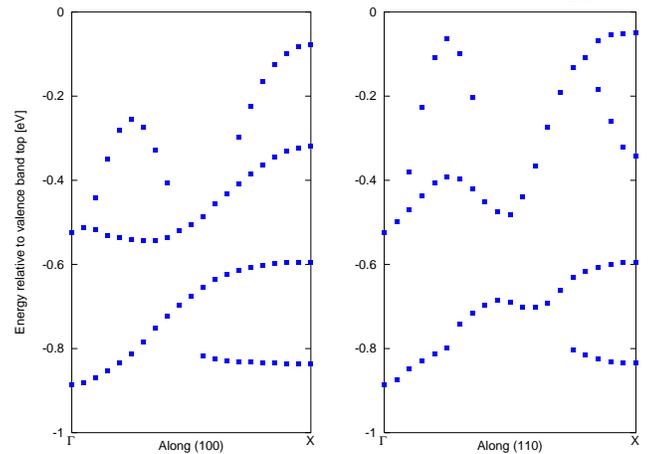}
\caption{\label{fig8} Fine structure of the
broad peak which forms the top of the valence band structure
in antiferromagnetic NiO. The symbols give the positions of
peaks with appreciable weight. }
\end{figure}
Finally, Figure \ref{fig8} gives the dispersion
of the `sub-bands' of the broad structure $A$ at the valence band edge.
This fine structure has not yet been resolved
experimentally as yet - however, Shen {\em et al.} found evidence
for at least three `subpeaks' and also for a quite substantial
dispersion, although this made itself felt only as a dispersion of
the line shape of the broad peak. Looking at Figures 9 and 10
of Ref. \cite{Shen_long} it would appear that along $(100)$ there is
an overall `upward' dispersion of the topmost peak 
$A$ as one moves from 
$\Gamma \rightarrow X$ with two local maxima of the upper edge of $A$
just after $\Gamma$ und just before $X$ with the whole
band complex being most narrow approximately halfway between 
$\Gamma$ and $X$. It can be seen already from Figure \ref{fig12}
that this dispersion of the peak-shape of $A$ is reproduced
quite well by theory.  Similarly, along
$(110)$ the brod band complex seems to have its
minimum width halfway between $\Gamma$ and $X$.
At least these qualitative results are quite
consistent with the dispersion in Figure \ref{fig8}.
Clearly a more detailed study of the fine structure of
feature $A$ would provide an 
interesting check of the present and other theories for the band
structure of NiO. Another stringent check for theory would
be to unravel the orbital character of the individual flat
bands such as $C$ and $D$ by studying their intensity as a function
of photon polarization and energy.\\
Finally, we mention the upper Hubbard band, with the corresponding
final states having predominantly $e^\dagger$-character.
The insulating gap has a magnitude of $4.3eV$, which is consistent
with experiment\cite{SawatzkyAllen}. 
Figure \ref{fig2} shows the angle integrated
(i.e.: $\bf{k}$-integrated) 
photoemission and inverse photoemission spectrum.
By and large there is reasonable agreement with experiment.
The fact that theory puts the
dispersionless bands $C$ and $D$ too close to the
top of the valence bands leads to a too weak
shoulder on the negative binding energy side
of the `main peak' at the top of the valence bands.
\begin{figure}
\includegraphics[width=\columnwidth]{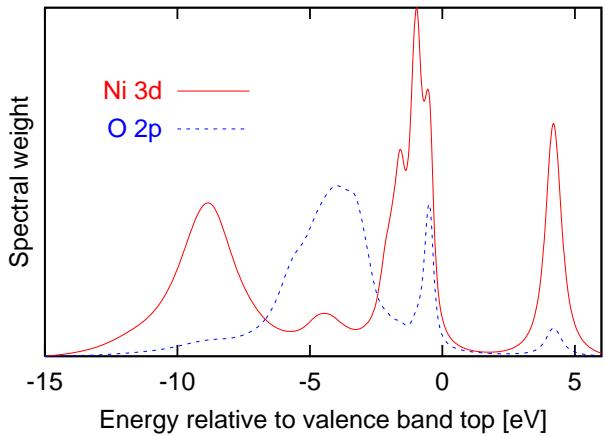}
\caption{\label{fig2} Momentum integrated
spectral weight for antiferromagnetic NiO.
To simulate a photoemisison spectrum the
Lorentzian broadening has been taken energy dependent according to
$\delta=0.4 eV + (\omega-1 eV)*0.1$.}
\end{figure}
\section{Conclusion}
In summary, we have presented a theory for the
single particle-excitations of charge transfer insulators.
The basic idea is to interpret the charge fluctuations
out of the purely ionic configuration as `effective Fermions'
and derive and solve an effective Hamiltonian for these.
This is the same physical idea which is
underlying both the Hubbard I approximation and the cluster
perturbation theory and, as demonstrated above,
when applied to a realistic model of a charge-transfer
insulator these methods, which so far have been restricted
to more `model-type' systems, give indeed quite satisfactory agreement
with experiment. The key approximation, namely to treat the
Hubbard-like operators describing the charge fluctuations
as free Fermions thereby is well justified, because of the
low density of these effective Fermions, which
renders their (strong) interaction largely irrelevant.
A systematical way to relax this approximation
would be the $T$-matrix approach, as
demonstrated by Kotov {\em et al}\cite{Oleg}.
It should be noted that the calculation is
computationally no more
demanding than a conventional band structure calculation and can
be `automated' almost completely. The weakest link in the chain
thereby is the necessity to perform an LCAO-fit to
an LDA band structure.\\
One important conceptual problem is the necessity to break the symmetry
which originates from the degeneracy of the ground state multiplet of
a single transition metal ion
and choose the `reference states' $|\Phi_{i,0}\rangle$ `by hand'.
However, one might as well consider choosing  an {\em ansatz} for these
reference states which takes the form of a linear combination
$|\Phi_{i,0}\rangle=\sum_\nu \alpha_{i\nu} |\nu\rangle$, where the sum
extends over the GS multiplet, and determine the
coefficients $\alpha_{i\nu}$ from the requirement of minimum total
energy. In this way spin and
orbital ordering could be studied in much the same way as lattice 
parameters are optimized in conventional LDA calculations
and since all `ingredients'
for the Goodenough-Kanamori rules\cite{Goodenough}
are taken into account, this
may be a quite promising method. Since spin-orbit coupling also can be 
trivially included in the exact diagonalization of the
isolated $d$-shells one might even hope to address magnetic 
anisotropies and/or
anisotropic exchange interactions.
A procedure for the improvement of CPT calculations on model
Hamiltonians which is similar
in spirit has been proposed by Potthoff {\em et al.}\cite{Chris}.\\
One major drawback of the theory
clearly is the approximate nature of the calculation of the spectral
weight. It should be noted, however, that there is a very
clear physical reason for this problem, namely the `compound nature'
of the ZRS-like states which form the top of the valence
band. If the present interpretation of these
states is the correct one, basically any theory will face similar problems.
One possible way out would be to derive a version of the
original CPT which can work with site-sharing clusters.\\
Finally, we would like to discuss the relationship between our
theory and previous workers in the field.
Manghi {\em et al.}\cite{Manghi} and 
Takahashi and Igarashi\cite{Igarashi} have calculated the
quasiparticle band structure of NiO more along the lines of
conventional many-body theory. Starting from a paramagnetic
LDA band structure (Ref. \cite{Manghi})
or an antiferromagnetic Hartree-Fock band structure
(Ref. \cite{Igarashi}) these authors added a self-energy constructed
within the local approximation to three-body scattering theory.
The obtained band structures show the same `large scale features'
as the one obtained here, but there are also
significant differences, particularly so near
the top of the valence band. More detailed comparison
with experiment seems necessary to discuss the merits
of the various theories.\\
Next, there is a clear analogy between the present theory
and the cluster method 
of Fujimori and Minami\cite{FujimoriMinami}
and va Elp {\em at al.}\cite{Elp}. With the exception
of $d^9\underline{L}^2$ states in the
photoemissions spectrum the present theory
employs the same type of
basis states as the cluster calculations. The only difference is,
that we designate one of the degenerate ground states
of $d^n$ as a `vacuum state' and interpret the other states
as `deviations' from this vacuum state. Those deviations
which carry the quantum number of an electron then are
considered as effective free Fermions. As discussed above,
the low density of these effective Fermions probably
make this a very good approximation.\\
There is also an obvious relationship between the present theory
and the work of 
Unger and Fulde\cite{UngerFulde}.
Using the projection technique developed by 
Becker and Fulde\cite{BeckerFulde}
these authors constructed an equation of motion for single-particle
spectral functions of the CuO$_2$ plane,
which is very similar to the ones which would be
obtained from our effective Hamiltonians. \\
Finally we address the work of
Bala {\em et al.}\cite{BalaOlesZaanen_I}, which is very similar in spirit
to the present theory. These authors derived a `Kondo-Heisenberg'-like
model operating in the subspace
of $d^8\underline{L}$ type states by eliminating - via canconical
transformation - the charge fluctuations between states of the
type $(^3A_{2g}d^8)\underline{L}$ and states of the type
 $d^7$ (their theory was concerned with the motion of a single hole
in an $O2p$ orbital).
Accordingly, their theory produced (in addition to the
free-electron-like $O2p$ bands) two weakly dispersive
bands - one for each of the `flavours' $e_g$ and $t_{2g}$
whereby the flavour stands for the symmetry of the linear
combination of $O2p$ orbitals around a given $Ni$ site.
Thereby Bala {\em at al.} actually went one step beyond the present
theory by taking into account the coupling of $O2p$-like holes
to the antiferromagnetic magnons - which is omitted in the
present theory.
In the present theory, no canonical transformation is
performed, so that also the high energy features
(satellite and upper Hubbard band) are reproduced.
Moreover, we also take the excited
multiplets of $d^8$ and their covalent mixing
with the $d^7$ multiplets into account, whence
we obtain a larger number of ZRS-like bands - consistent with
experiment. Experimentally the impact of the coupling to magnons
which is ignored in the present theory but treated
accurately in the work of Bala {\em at a.}
could be studied only by considering the `fine structure'
of the broad peak at the valence band top. These states
seem to have an appreciable dispersion which might or might not
be influenced by the coupling to magnons.

\end{document}